 \documentclass[prl,aps,twocolumn,psfig,preprintnumbers,amsmath,
amssymb,showpacs]{revtex4}
\usepackage{graphicx}\usepackage{epsfig}
\usepackage{amsmath}\usepackage{amssymb}\usepackage{amsfonts}
\usepackage{bm}
\usepackage{exscale}
\usepackage{dcolumn}    
\usepackage{color}
\usepackage{soul}
\usepackage{ulem}
\begin{document}
%
%
\title{Pairing effects on spinodal decomposition of asymmetric nuclear matter}
%
%
\author{S.Burrello$^{1}$}
\author{M.Colonna$^{1}$}
\author{F.Matera$^{2}$}
\affiliation{$^{1}$~INFN-LNS, Laboratori Nazionali del Sud, 95123 Catania, Italy \\
$^{2}$~Dipartimento di Fisica e Astronomia, and Istituto Nazionale di Fisica Nucleare, 50019 Sesto Fiorentino, Firenze, Italy}

%
%

%
%
\begin{abstract}
We investigate the impact of pairing correlations on the behavior of
unstable asymmetric nuclear matter at low temperature. We focus on the
relative role of the pairing interaction, coupling nucleons of the same
type (neutrons or protons), with respect to the symmetry potential, which
enhances the neutron-proton attraction, along the clusterization process
driven by spinodal instabilities.
It is found that, especially at the transition temperature from the
normal to the superfluid phase, pairing effects may induce significant
variations in the isotopic content of the clusterized matter.
This analysis 
is potentially useful to gather information on
the temperature dependence of nuclear pairing and, in general, on the 
properties of clusterised low-density matter, of interest also in the
astrophysical context.
\end{abstract}
%
%
\maketitle
%
%
%
%
%
%
%


%
Investigations of many-body interacting systems have always
been an exciting and attractive field in different domains of
physics. Indeed understanding the properties of complex systems 
in terms of their constituent particles and the interaction among them
is a true challenge.   
The original (and unsolvable) quantal  many-body problem, is often approached
adopting the mean-field approximation, yielding a so-called 
effective interaction \cite{Rein,Bend}. %
However, suitable extensions of mean-field models have been introduced to take explicitly
into account the effects of relevant interparticle correlations. 
This is the case, for instance, of pairing correlations which occur, under
suitable conditions, in fermionic systems \cite{Sch64}.

Many efforts are currently focused on the study of 
the properties of complex nuclei and infinite nuclear matter.
It is well known that nucleons can form paired states, analogous to the way electrons pair in metals, yielding 
a superconducting phase \cite{Sch64}. Pairing effects  
on nuclear masses are widely investigated nowadays, also in connection with
astrophysical applications requiring the knowledge of the mass of very neutron-rich nuclei, 
which play a crucial role in the $r$ process of nucleosynthesis \cite{Arn07}.  
Moreover, the presence of neutron superfluidity in the crust and the inner part of neutron stars
is considered well established in the physics of these compact stellar objects,
and has a significant effect on cooling processes \cite{Pa09} 
and glitch phenomena \cite{Ca00,Se05}.

Complex many-body systems are also characterized by 
the possible occurence of different kinds of phase transitions. 
For nuclear matter at sub-saturation density and relatively low temperature ($T \lesssim 15 MeV$)
liquid-gas phase transitions are expected to appear, driven by the
unstable mean-field.  Such a process is closely linked to the multifragmentation mechanism
experimentally observed in nuclear reactions \cite{Borderie}
and to the occurrence of clustering phenomena in the inner crust of neutron stars
\cite{Lattimer,Camille}.  
Owing to the two-component structure of nuclear matter, 
a central role 
in this mechanism pertains to the 
density behavior of the isovector part of the effective interaction 
and the corresponding term in the 
nuclear Equation of State, the symmetry energy \cite{rep,rep1},
on which many investigations are concentrated \cite{betty,shetty,galichet,Chimera,Bao}.      
Indeed, as also pointed out for the pairing interaction, 
this information is essential 
for the understanding of nuclear structure \cite{Colo,virgil_pygmy} and of the properties of neutron stars 
\cite{Lattimer,Duc}.
Along a phase separation process, the symmetry energy influences significantly the so-called
isospin distillation mechanism, which leads to a different species concentration in
the two phases, namely a more symmetric (with respect to the initial system)
liquid phase and a more asymmetric gas phase 
\cite{rep1,Col_PRL}. 

The aim of this Letter is to evaluate the impact of pairing correlations,
which are mostly active at low density, on mechanical (spinodal) instabilities
of asymmetric nuclear matter. 
Since pairing between protons and neutrons is strongly quenched in asymmetric matter 
\cite{Lom00}, here we consider a pairing interaction acting only between identical nucleons in a 
spin--singlet state.

Our  study is undertaken in the framework of the Hartree-Fock-Bogolyubov approach, which includes
 in a unified formalism the pairing and the mean-field effective interactions \cite{Bend}. 
As far as the pairing interaction is concerned, we adopt a local interaction of strength $v_{\pi,q}$, where $q = n,p$ 
denotes neutrons
or protons respectively.



Though the density dependence of the pairing force is still poorly known, it is often assumed that 
$ v_{\pi q}$ depends only on the density $\rho_q$ of the considered species \cite{Be91,Chamel} , 
thus we write: 

\begin{equation}
\label{eq:strength}
v_{\pi q} (\rho_n, \rho_p) \equiv v_\pi (\rho_q) = V_\pi \biggl [ 1 - \eta \biggl (\frac {2\rho_q}{\rho_{0}} \biggr )^\alpha \biggr ],
\end{equation}
where 
$\rho_0$ is the nuclear saturation density. 
The parameters in Eq.(\ref{eq:strength}), $V_\pi $, $\eta$ and $\alpha$, are usually fitted directly to experimental data, taking as reference the pairing gap in finite nuclei.
However, in order to establish a
connection with realistic nucleon-nucleon forces, it has been recently proposed to determine the parameters of the pairing
interaction 
by fitting the $^1S_0$ paring gaps in 
infinite nuclear matter as obtained 
in the BCS approximation \cite{Garr} or in Brueckner calculations \cite{Lomb}. 
Then the density dependence of the pairing strength can be calculated exactly, for any given  $^1S_0$ pairing-gap function $\Delta(\rho_q)$
by inverting the
gap equation written below:


\begin{equation}
I_\Delta \equiv - v_{\pi} (\rho_q) \int \frac {d \mathbf p} {(2 \pi \hbar)^3} \frac {1}{2\xi} \Bigl [ 1 - 2 F_q(\mathbf p) 
\Bigr ] = 1,
\end{equation} 
where 
\begin{equation}
\label{eq:f_upup}
F_q(\mathbf p) 
= \frac {1} {2} \left [ 1 - \frac {\xi} {E_\Delta} \tanh \left ( \frac {\beta E_\Delta}{2} \right ) \right ]
\end{equation}
represents the particle occupation number.

In the previous equations  $\beta$ indicates the inverse of the temperature $T$ and 
$E_\Delta = \sqrt{\xi^2 + \Delta^2}$, where  $\xi =  \left ( \frac {\mathbf {p}^2} {2m} 
- \mu^*_q \right )$, being $m$ the nucleon mass.  The reduced chemical potential 
$\mu^*_q$ 
can be obtained 
by fixing the 
particle number density: 

\begin{equation}
I_\rho \equiv  
\int \frac {d \mathbf p} {(2 \pi \hbar)^3}\left [ 1 - \frac {\xi} {E_\Delta} \tanh \left ( \frac {\beta E_\Delta}{2} 
\right ) \right ] = \rho_q\label{eq:densità}.
\end{equation}
It may be useful to recall that,
in absence of pairing, 
the reduced chemical potential $\mu^*_q$ 
coincides with that of a non interacting Fermi gas, $\mu^*_{q,F}$ .  
It should be noticed that, 
owing to the zero range of the pairing interaction, 
a cutoff has to be introduced in the gap equations to avoid divergences (see for instance \cite{Dug}). 
Here we adopt the energy cutoff $\epsilon_\Lambda = 16~MeV$ \cite{Chamel}.

We will take as reference the pairing gap in pure neutron matter at zero temperature,  
as obtained in Brueckner calculations employing the realistic Argonne $v_{14}$ potential (see Ref.\cite{Lomb}).
The corresponding values 
are plotted in Fig.1 (black plus symbols) as a 
function of the neutron density $\rho_n$.  
The maximum value of the gap ($\Delta \approx 3~MeV$) is reached at the density $\rho_M = 0.02~fm^{-3}$. 
Knowing $\Delta$, 
from Eqs. (2-4)  it is possible to extract the pairing interaction $v_\pi$ as
a function of the neutron matter density.
The results obtained can be fitted by the functional dependence expressed by Eq.(1), with the 
following parameters: $V_\pi = -1157.51~MeV$, $\eta = 0.884$, $\alpha = 0.256$ and 
the corresponding behavior is shown in the bottom panel of Fig.1.
Once the strength of the pairing interaction is fixed, Eqs.(2-4) allow one to evaluate
the pairing gap at finite temperature $T$.
Results are shown in Fig.1, top panel. 
The critical temperature ($T_c(\rho_q)$) of the transition to the superfluid/superconducting phase
(where the energy gap $\Delta$ starts to appear)
depends on the density considered and is equal to
$T_c = 1.8 ~MeV$ at $\rho_M$, see the inset in Fig.1, top panel.
The results discussed above are extended to the $pp$ case, assuming that the pairing strength is the same as in the $nn$ case,
just depending on the density of the species considered.

\begin{figure}
\includegraphics[width=8.0cm]{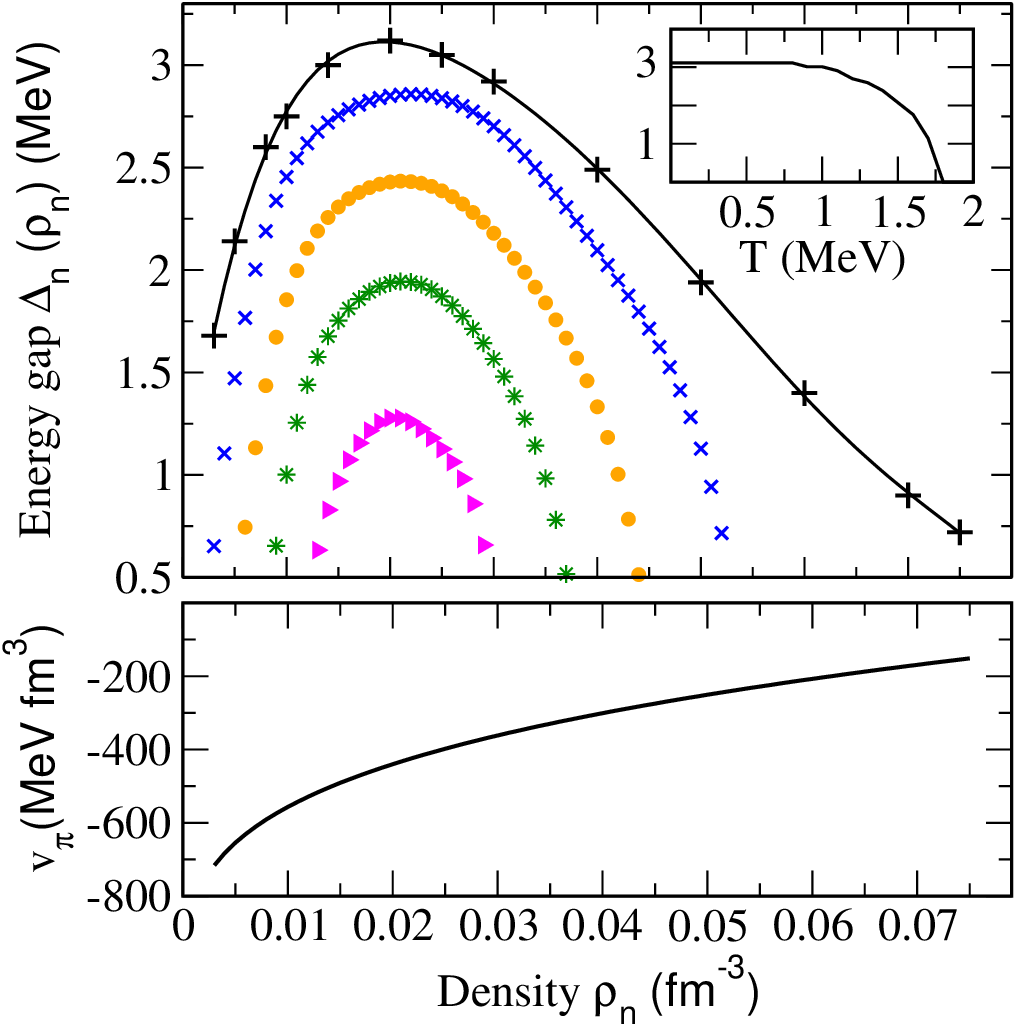}

\caption{(Color online) 
Top panel:
The energy gap, as obtained in Brueckner calculations
of pure neutron matter at zero temperature, as a function of the neutron density (black plus symbols). 
The figure also shows the energy gap obtained, solving Eqs.(2-4), at
several $T$ values: 1 MeV (crosses), 1.3 MeV (dots), 1.5 MeV (stars), 
1.65 MeV (triangles). The inset displays the energy gap as a function of
the temperature, for the density $\rho_M = 0.02~fm^{-3}$.
Bottom panel:
The strength of the pairing interaction $v_\pi$ as a function of the neutron
matter density. }
\end{figure}

Adopting a simplified  Skyrme-like effective interaction 
for the mean-field \cite{BaranNPA703}, 
the 
nuclear energy density functional can be written as it follows: 

\begin{align}
\rho \frac {E}{A} =& 2\sum_q \int \frac {d \mathbf p}{(2 \pi\hbar)^3} F_q(\mathbf p)  \frac {\mathbf {p}^2} {2m} 
+ \frac{1}{4} \sum_q v_\pi (\rho_q)\tilde{\rho}_q^*\tilde{\rho}_q  \notag\\
      &+ \rho \biggl [ \frac {\mathcal{A}} {2} \biggl (\frac {\rho}{\rho_0} \biggr ) + \frac{\mathcal{B}}{\sigma +1} \biggl (\frac {\rho}{\rho_0}\biggr )^\sigma 
+ \mathcal{C}^{pot}_{sym}(\rho)  I^2  \biggr ]  \label{eq:ea_pairing},
\end{align}
where $\rho = \rho_n + \rho_p$, $I = \frac {(\rho_n - \rho_p)} {\rho} $ is the asymmetry parameter and 
$\tilde{\rho}_q = 2 \Delta(\rho_q)/ v_\pi(\rho_q)$ denotes the so-called anomalous density. 
The coefficients $A=-356~ MeV$,$B=303~ MeV$ and the exponent $\sigma=\frac{7}{6}$, 
characterizing the isoscalar part of the mean-field, are fixed requiring 
that the saturation properties of symmetric 
nuclear matter, 
with a compressibility of $200~MeV$, are reproduced. 
As far as the isovector part of the nuclear interaction is concerned, 
we consider, as a possible description of 
the local density  ($\rho$) dependence of  
the symmetry energy coefficient $C_{sym}^{pot}$,
two representative parametrizations: 
one with a linearly increasing behaviour   
 with density (asy-stiff),  $C_{sym}^{pot}(\rho) = 112.5~\rho$ (MeV),
and one which saturates 
above normal
density (asy-soft),  $\displaystyle{C_{sym}^{pot}(\rho)}=\rho~(241-819~\rho)$ (MeV) 
\cite{BaranNPA703,rep1}. 

The mean-field potential $U_q$ can be derived from the potential part of the energy density functional, Eq.(5):
${U}_q = \left [ \frac {\partial (\rho E_\textup{pot} /A)} {\partial \rho_q} \right ]_{\tilde{\rho_q}}$ 
\cite{Cha08}.
We notice that, because of the density dependence of the pairing strength, 
$U_q$ gets a contribution also from the pairing energy density (second term in the r.h.s. of 
Eq.(5)): $ U_q^\pi = \Delta^2/{v^2_\pi}~ \partial v_\pi/\partial \rho_q $.

We now turn to examine the possible occurrence of mechanical (spinodal) instabilities in 
asymmetric nuclear matter at a fixed temperature $T$. 
The spinodal region is defined as the ensemble of unstable points, for
which the free-energy surface is concave in at least one direction. 
This is determined by the curvature matrix:
\begin{equation}
 C = \begin{pmatrix}
a                 & c/2 \\
c/2 & b
\end{pmatrix},
\end{equation}
where $a = \partial \mu_p / \partial \rho_p$, $b = \partial \mu_n / \partial \rho_n$ 
and  $c = 2\partial \mu_p / \partial \rho_n$. $\mu_q = \mu^*_q + U_q$ denotes
the chemical potential.  
The lower eigenvalue gives the minimal free-energy curvature. 
If the latter is negative, the associated eigen-vector  gives the direction
of phase separation \cite{rep,rep1}. 



In asymmetric nuclear matter ($\rho_n \neq \rho_p$) 
at low density, 
instabilities correspond to isoscalar-like density oscillations: The two species move in phase but 
with different amplitudes, according to the eigen-vector components, $( \delta\rho_p, \delta\rho_n ) $.
In particular, defining the angle $\theta$ as $\tan \theta 
= \delta\rho_n / \delta\rho_p$, from the
diagonalization of the $C$ matrix one obtains \cite{Virgil_PRL}: 
\begin{equation}
\label{eq:tan_2}
\tan 2 \theta = \frac {c}{a - b}.
\end{equation}
It is generally observed that the asymmetry of the 
instability direction,  
$\delta I = (\delta\rho_n - \delta\rho_p)/(\delta\rho_n + \delta\rho_p)$,
is smaller than the system initial asymmetry, leading to the formation of more symmetric nuclear clusters. 
This is the so-called isospin distillation mechanism, that is mainly ruled by the effect of the symmetry potential,
which enhances the neutron-proton attraction.  Indeed, a stiffer symmetry energy tends to reduce the difference between 
proton and neutron chemical potential derivatives, i.e. the term $(a-b)$ in Eq.(7), and the angle $\theta$ gets
closer to $45^0$. 
Thus neutrons and protons 
oscillate with close amplitudes, in spite of the system initial asymmetry. 
  
Here our aim is to investigate how pairing correlations may affect these features.
Indeed, as seen in Fig.1, the pairing energy gap is maximum at neutron and/or proton low
densities compatible with the nuclear spinodal zone ($\rho < 0.1~fm^{-3}$ at $T$ = 0).
The strength of the instability, i.e. the amplitude of the negative eigenvalue of $C$,  
is mainly determined by the isoscalar part of the nuclear mean
field potential which is by far the dominant term of the interaction. 
Thus the pairing interaction has practically no effect on it. 
On the other hand, the distillation mechanism, which is connected to the strength of the symmetry 
potential, 
may be affected by the paring interaction, that couples nucleons of the
same type. 
 
To undertake this analysis we need to evaluate the expression of the elements of the $C$ matrix 
in presence of pairing correlations. We notice that, for the pairing interaction considered here,
the term $c$ gets no contribution from the
pairing interaction.
The calculation of $a$ and $b$ requires the knowledge of the following derivatives:
$\partial\mu^*_q / \partial \rho_q$ and  $\partial \Delta / \partial \rho_q$
(the latter appears in the derivative of the potential $U^\pi_q$), 
which can be obtained solving the following set of equations, derived from Eqs.(2) and~(4):
\begin{equation}
\begin{cases}
\displaystyle \frac{\partial I_\Delta}{\partial \rho_q} 
+& \displaystyle \frac{\partial I_\Delta}{\partial \mu^*_q}\displaystyle \frac {\partial \mu^*_q}{\partial \rho_q} 
+ \displaystyle \frac{\partial I_\Delta}{\partial \Delta} 
\displaystyle \frac{\partial \Delta}{\partial \rho_q}  = 0 \\
& \displaystyle \frac{\partial I_\rho}{\partial \mu^*_q} \displaystyle\frac {\partial \mu^*_q}{\partial \rho_q} 
+ \displaystyle \frac{\partial I_\rho}{\partial \Delta}  
\displaystyle \frac{\partial \Delta}{\partial \rho_q} = 1 
\end{cases}
\end{equation}
One relevant quantity, to evaluate the width of the isospin distillation effect, is
the difference $\gamma = (a - b)$
of Eq.(7). 
Thus we consider its percentage variation in 
superfluid/superconducting nuclear matter, with
respect to normal nuclear matter, at different global densities and asymmetries,
and zero temperature.  
Results are displayed in Fig.2, in the case of the asy-stiff symmetry potential.
One can see that it is possible to reach an effect of $20\%$ 
for the variation
of $\gamma$, at total densities around 
$0.08~fm^{-3}$ and $I\approx 0.1-0.2$.  
\begin{figure}
\includegraphics[width=8.0cm]{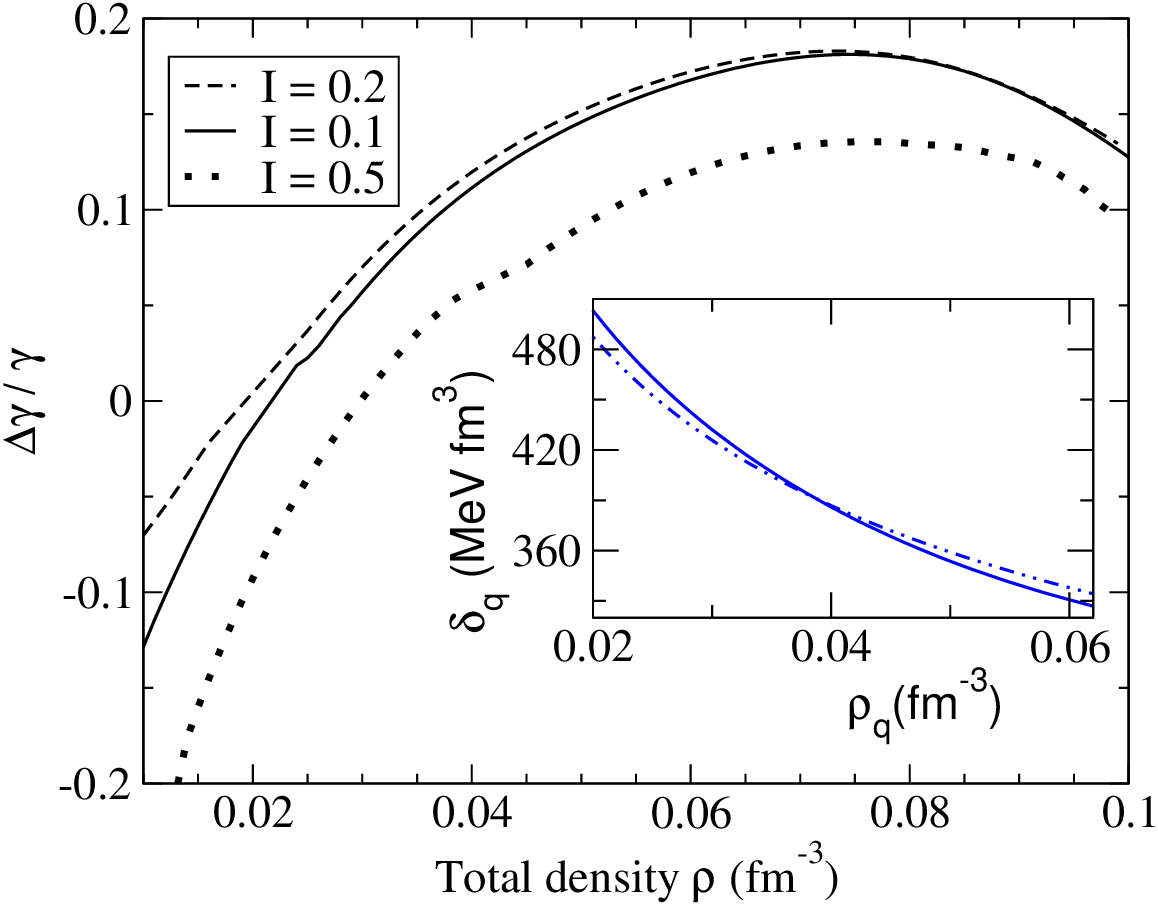}
\caption{(Color online)
The percentage variation of the quantity $\gamma$ (see text) as a function of the
matter density, at zero temperature and for three values of the asymmetry I.
The inset shows the density
behavior of the quantity $\delta_q$ (see text), in normal (dashed line) and
superfluid (full line) matter.
}
\end{figure}
We recall that the features of the unstable modes are determined by 
derivatives of proton/neutron chemical potentials, thus  
a deeper insight into the amplitude of the pairing effect can be obtained by looking directly
at the quantity $\delta_q = \partial\mu^*_q / \partial \rho_q + \partial U^\pi_q/ \partial \rho_q$.
The latter is displayed in the inset of Fig.2, 
as a function of the
density $\rho_q$, together with the corresponding values of normal nuclear matter
(which are nothing but the density derivative of $\mu^*_{q,F}$). 
The two curves exhibit a different slope mostly around
$\rho_q \approx 0.04 fm^{-3}$, thus pointing to the 
region of total density ($\rho \approx 2\rho_q = 0.08 fm^{-3}$) where one expects
the largest influence of pairing correlations on the difference $(a-b)$, as already evidenced by the results 
shown in Fig.2. 

Let us move to consider nuclear matter at finite temperature $T$.  
The quantity $\delta_q$ is represented in Fig.3, as a function of $T$, 
for three values of the density $\rho_q$ 
($\rho_M/2$, $\rho_M$, $2\rho_M$).
Calculations including pairing correlations (blue circles) are compared with 
the results of normal nuclear matter (red dashed lines). 
As a rather interesting effect,
we clearly observe the appearance of discontinuities, at the critical temperature, in the behavior of~$\delta_q$. 

It is well known that at the critical temperature $T_c(\rho_q)$ 
the heat capacity exhibits a discontinuity \cite{Sch64}. 
Hence, together with this widely discussed effect, 
one has to notice that discontinuities also appear
in the behavior of the density derivative of the chemical
potential, $\partial \mu_q / \partial \rho_q$, which is connected to the matter compressibility. 
It is also rather interesting to observe that, owing to the features of 
the gap function in nuclear matter (see Fig.1), the jump
disappears at the density $\rho_q = \rho_M$ (panel (b)), where the 
energy gap $\Delta$ is maximum at all temperatures.

\begin{figure}
\includegraphics[width=8.0cm]{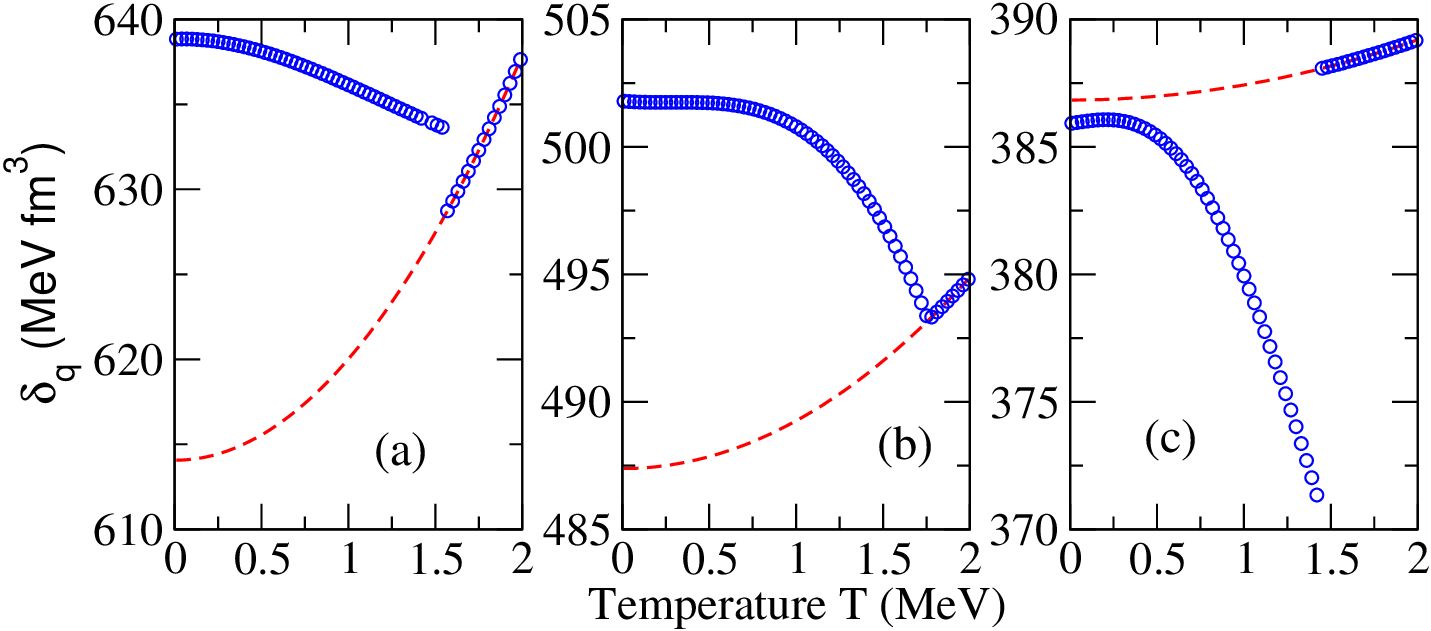}

\caption{(Color online)
The quantity $\delta_q$ (see text) is represented as a function of the temperature,
for three $\rho_q$ density values: $0.5~\rho_M $ (a), $\rho_M$ (b) and $2~ \rho_M$
(c). Circles indicate the calculations including pairing effects, whereas dashed
lines are for normal nuclear matter. 
}
\end{figure}

Finally we turn to discuss the impact of pairing correlations
directly on the asymmetry $\delta I$ 
of the instability direction. 
Results are displayed in Fig.4, 
for nuclear matter at $\rho = 0.08~fm^{-3}$, where the largest effects have been
observed at zero temperature (see Fig.2),  and three values of the asymmetry $I$. 
The two parametrizations of the symmetry energy introduced above are considered. 
First of all we notice that the overall effect of the isospin distillation
mechanism is rather important.  Indeed $\delta I / I $ is quite lower than $1$ in all cases. 
The effect is larger in the asy-stiff case, which is characterized by a steeper variation
of the symmetry energy with the density, 
in agreement with previous studies \cite{rep1,Matera}.   
The asymmetry $\delta I$ is more sensitive to the symmetry energy
parametrization (black vs. red lines) than to the introduction
of pairing correlations (full vs. dashed lines). 
Thus our results essentially confirm the leading role of
the symmetry energy  in the isospin distillation mechanism.
However, new interesting effects appear at moderate temperatures. 
Owing to the trend followed by the chemical potential derivatives
(see Fig.3),
the calculations including the pairing interaction exhibit 
two discontinuities, in correspondence of 
neutron and proton critical temperatures, which may cause significant variations of   $\delta I$.  
As observed in Fig.3, the $\delta_q$ discontinuity is more pronounced at densities
above $\rho_M$, around $\rho_q \approx 0.04~fm^{-3}$. This explains why
the largest effect for $\delta I$ are seen at  
small asymmetries (see panel (a)), 
i.e. for neutron and proton densities $\rho_q$ close to $2\rho_M$. 
Hence, under suitable density and  temperature conditions, 
pairing correlations may lead to significant deviations of the asymmetry from its average. 

To conclude, 
\begin{figure}
\includegraphics[width=8.0cm]{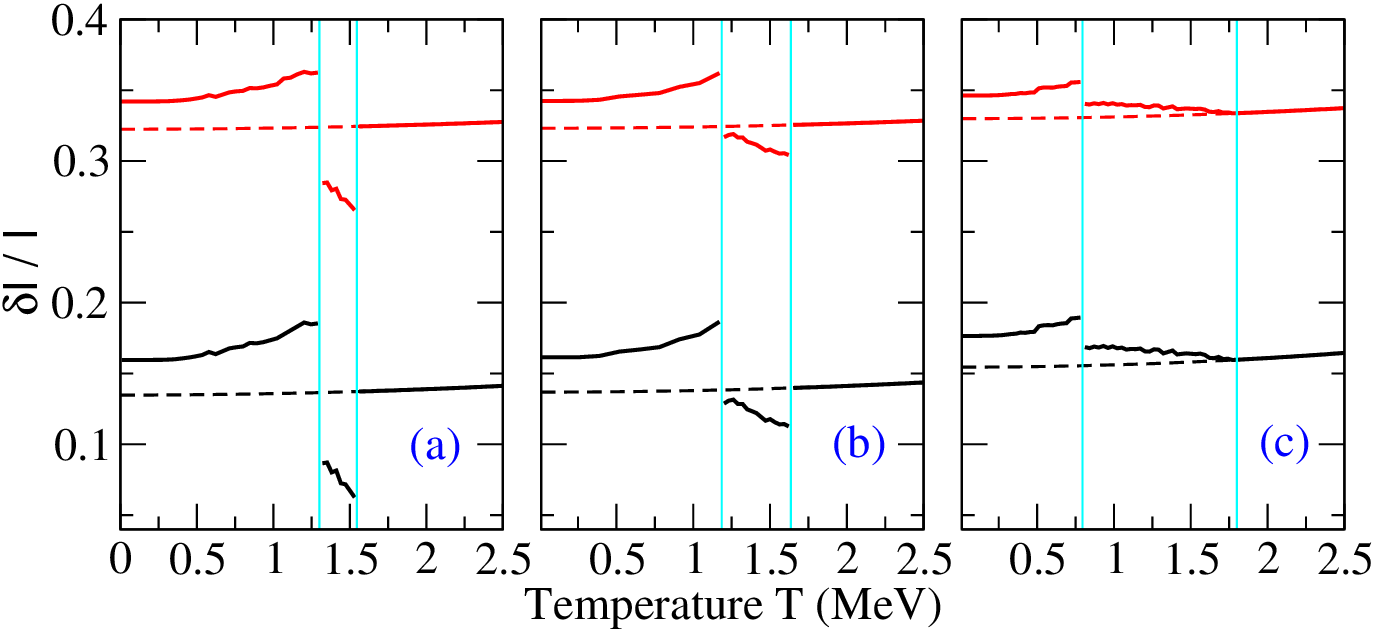}  
\caption{(Color online)
Full lines: The asymmetry of the unstable oscillations is plotted as a function of the temperature, 
for nuclear matter at total density $\rho = 0.08~fm^{-3}$ and three asymmetry
values: I = 0.1 (a), 0.2 (b), 0.5 (c). 
Results are shown for the asy-stiff (black) and the asy-soft (red) parametrizations. 
Dashed line: the same quantity but for normal nuclear matter. 
In each panel the two vertical lines indicate the position of neutron and proton 
critical temperatures.}
\end{figure} 
we have presented an analysis framed in the general context 
of  two-component fermionic systems feeling pairing correlations.  
We have focused on the interplay 
between the pairing force, coupling particles of the same type, 
and the isovector interaction, which on the contrary enhances the attraction between 
particles of different kind. 
This study has been conducted for unstable asymmetric nuclear
matter at low temperature, where it is shown that pairing correlations may have non negligible
effects, especially around the transition temperature to the superfluid/superconducting phase,  on the isotopic features of the density fluctuations leading to cluster
formation. 
These results are relevant to the study of pairing correlations in 
low temperature nuclear fragmentation processes, 
as far as effects related to level density and isotopic composition of the
primary fragments are concerned \cite{Michela}. 
Moreover, the presence of pairing correlations may also affect the description
of low-density clustering phenomena occurring in the crust of neutron stars
\cite{Camille,pastas}.  

{\bf Acknowledgments} - 
Inspiring discussions with U.Lombardo are gratefully acknowledged.

\end{document}